# A Co-Simulation Study to Assess the Impacts of Connected and Autonomous Vehicles on Traffic Flow Stability during Hurricane Evacuation

**Zaheen E Muktadi Syed**
Department of Civil, Environmental and Construction Engineering
University of Central Florida
Email: zaheensyed@ucf.edu

**Samiul Hasan, PhD**
(Corresponding author)
Associate Professor
Department of Civil, Environmental and Construction Engineering
University of Central Florida
Email: samiul.hasan@ucf.edu

**ABSTRACT**
Hurricane evacuation has become a major problem for the coastal residents of the United States. Devastating hurricanes have threatened the lives and infrastructure of coastal communities and caused billions of dollars in damage. There is a need for better traffic management strategies to improve the safety and mobility of evacuation traffic. In this study, hurricane evacuation traffic was simulated using SUMO, a microscopic traffic simulation model. The effects of Connected and Autonomous Vehicles (CAVs) and Autonomous Vehicles (AVs) were evaluated using two approaches: (i) using the state-of-the-art car-following models available in SUMO and (ii) a co-simulation study by integrating the microscopic traffic simulation model with a separate communication simulator to find the realistic effect of CAVs on evacuation traffic. A road network of I-75 in Florida was created to represent real-world evacuation traffic observed in Hurricane Irma's evacuation periods. Simulation experiments were performed by creating mixed traffic scenarios with 25%, 50%, 75%, and 100% of different vehicle technologies including CAVs or AVs and human-driven vehicles (HDV). Simulation results suggest that the CACC car-following model, implemented in SUMO and commonly used in the literature to represent CAVs, produces highly unstable results. On the other hand, the ACC car following model, used to represent AVs, produces better and more stable results. However, in a co-simulation study, to evaluate the effect of CAVs in the same evacuation traffic scenario, results indicate that with 25% CAVs, the number of potential collisions decrease up to 42.5%.

**Keywords:** Hurricane Evacuation, Connected and Autonomous Vehicles (CAVs), SUMO, Omnet++

Syed and HasanINTRODUCTION

Hurricanes have become more common and intense in the coastal regions of the United States. Evacuations are typically ordered to save lives during major hurricanes. For example, during Hurricane Irma approximately 6.5 million people from the coastal regions of Florida were asked to evacuate. Since there were only two highways, a massive traffic congestion and a high number of crashes were observed in those highways during the evacuation period. A high influx of traffic results in unstable traffic flows and congestion. Previous studies found that traffic flow significantly varies during evacuation with highly fluctuating speeds which may lead to rear end collisions (*1*) (*2*).

Although various traffic management strategies such as contraflow and use of emergency shoulder helped accommodate high volumes of evacuation traffic, potential safety issues of evacuation traffic are of major concerns. For instance, Rahman et. al (*3*) found that for a high volume of traffic at an upstream location and a high variation of speed at a downstream location, the likelihood of a crash increases during evacuation. Stabilizing traffic flow by reducing stop and go movements can help improve evacuation traffic safety. In our previous study (*4*), we found that by equipping 25% vehicles with Adaptive Cruise Control (ACC) technology, the number of potential conflicts can be reduced by 49.7%. Emerging vehicular technologies such as Connected and Autonomous Vehicles (CAVs) have shown further promises in stabilizing traffic. Field experiments have shown that a controller deployed in a single autonomous vehicle out of twenty vehicles in a circular track can stabilize traffic and reduce congestion (*5*).

As smart vehicular technologies are increasingly adopted, it is imperative that these technologies are evaluated for real-world traffic scenarios that are more complex and have higher likelihood of crashes. In this study, we evaluated the performance of Autonomous Vehicles (AVs), Connected and Autonomous Vehicles (CAVs) and human-driven vehicle (HDVs) in a simulation environment (SUMO) that was calibrated to represent an evacuating traffic for Hurricane Irma in Florida. The safety impact of each technology was measured using different surrogate safety measures. Simulation experiments were designed to gain insights on how evacuation traffic safety and stability would change for different proportions of AV/CAV technology to co-exist in a network. The simulation experiments were carried in two different simulation environments: (i) adopting common car following models using SUMO only and (ii) integrating a communication simulator (Omnet++) and a traffic simulator (SUMO) to simulate the effects of CAVs on evacuation traffic safety.

This study contributes to literature by integrating a communication simulator and a traffic simulator-- enabling a more realistic assessment of the effects of vehicular connectivity. Our results indicate that CAVs simulated by a conventional car-following model may not fully capture the benefits of vehicular connectivity in such evacuation scenarios. This insight emphasizes the significance of representing communication capabilities in a realistic way when modeling CAVs, ensuring a more comprehensive understanding of their potential impact in complex traffic scenarios.

## Literature Review

### Review of the Car Following Models for HDVs, AVs and CAVs

In a simulation environment, different car following models are used to represent driving behavior. Some notable car following models are Gipps' model (*6*), the Krauss model (*7*), Intelligent Driver Model (IDM) (*8*), and Wiedemann car-following model (*9*). Previous research shows that the Krauss car following model can be used to represent human-driven vehicles (HDVs) due to its ability to emulate naturalistic driving behavior of humans and has been found to be consistent with the real-world traffic patterns found from detector data (*4, 10, 11*). The Krauss model also follows the speed change of the leader vehicle and have less errors in speed predictions for unsteady traffic condition than other car following models like IDM (*12*).

The adaptive cruise controller (ACC) has been increasingly available in new cars and proved to increase traffic throughput and safety. Relying on onboard sensors only, ACCs are state-of-the-art driving assistance systems which maintain a constant headway between subject and leader vehicle by adjusting the subject vehicle's speed and acceleration. The Cooperative Adaptive Cruise Controller (CACC) is an emerging technology that incorporates vehicle to vehicle (V2V) or vehicle to infrastructure (V2X) communications in addition to ACC sensors to make a subject vehicle's speed change decisions. CACC uses the same maneuvers as ACC but additionally enjoys the benefits of data available through vehicular communication. Previous simulation studies used the ACC car-following model to represent autonomous vehicles (AV) and the CACC car following model to represent Connected Autonomous Vehicles (CAV). Milanes et al. 2014 (*13*) developed the ACC and CACC car following models based on mathematical derivations to reflect the car following characteristics of the ACC and CACC equipped vehicles and on the data collected from field tests. The model was further improved by Xiao et al., 2017 (*14*) and Liu & H., 2018 (*15*).

**A Review of the Past Micro-simulation Studies**

Previously, many studies run simulation models to assess the effects of connected autonomous vehicle (CAVs) and autonomous vehicles (AVs) for different traffic network and conditions. The results of a simulation experiment highly depend on the assumptions made during experimental setups. Arvin et al. (*16*) evaluated ACC and CACC car-following models at an intersection, showing that 25% ACC-equipped vehicles led to marginal safety improvements, with better impacts observed at 40% ACC penetration. However, it is important to note that the study lacked well-defined vehicle parameters, such as the desired speed and headway of ACC/CACC-equipped vehicles and was limited to a single intersection. Guériau & Dusparic (*11*) studied various mixed proportions of autonomy using different road networks and CACC car-following model, indicating that CAVs enhanced traffic efficiency but were dependent on network characteristics and congestion levels. However, they pointed out the limitations of using the Intelligent Driver Model (IDM) for fully autonomous vehicles. Two additional simulation-based studies (*10*) (*17*) highlighted the positive effects of CAVs on traffic flow and reduced potential traffic conflicts but did not consider highly congested freeway scenarios observed during hurricane evacuations.

This literature review synthesizes evidence demonstrating the potential benefits of CAV adoption in enhancing traffic safety and efficiency. However, specific considerations such as traffic network types and autonomous vehicle modeling specially during complex traffic situations such as hurricane evacuation would require further exploration.

**A review of co-simulations studies**

Vehicular Ad-hoc Networks (VANETs) are integral to Connected and Autonomous Vehicles (CAVs) as they facilitate information exchange between vehicles and infrastructure through Dedicated Short-Range Communication (DSRC). However, practical scenarios introduce challenges like packet losses and message reception delays, impacting communication reliability (*18*). To address this, hybrid frameworks coupling traffic and communication simulators have been developed (*19–21*). While previous studies (*4, 10, 11*) investigated influence of CAVs on traffic safety and efficiency at various penetration rates, they often overlook realistic communication protocols, large reaction times, vehicle modeling, and traffic scenarios. Few studies investigated the effect of imperfect communication links on traffic performance. Recent research with hybrid frameworks demonstrates that CAVs can enhance traffic efficiency in congested scenarios, particularly at high penetration rates. However, the impact varies based on communication

__Syed and Hasan

reliability (*22, 23*). Also, the effect of CAVs in complex traffic situations like hurricane evacuations remains understudied. This study utilizes the VEINS framework (*20*), integrating SUMO and OMNET++, to comprehensively assess the impact of CAVs during hurricane evacuations for different market penetration rates, aiming to fill this research gap.

**METHODS**

In this study, we simulated a portion of a freeway network used for hurricane evacuation after calibrating a simulation model so that it represents the traffic condition observed during evacuation. We used a microscopic traffic simulator called Simulation of Urban Mobility (SUMO) version 1.10.0. For simulation, we adopted the traffic network used in Rahman et al., 2021 (*4*). However, we made a few modifications in that network to make it more realistic.

**Network Design**

The simulated traffic network consists of a 9.5-miles long road segment in I-75 between Ocala and Gainesville, Florida (as shown in Figure 1) with two entry and two exit ramps. The traffic data used in this study were obtained from the Regional Integrated Transportation Information System (RITIS) platform (*24*) for the selected road segments. The main modifications were made in the ramp merging sections. Intuitively, vehicles tend to change lanes as soon as they merge into the freeway rather than waiting for the end. So, a suggested approach is to elongate the speeding section in simulations such that vehicle have more time to change the lane and prevent any unlikely queue formation at the merging point. Earlier, in our simulation, we observed that when vehicles were merging from a ramp to the freeway or from the freeway to an exit ramp, some vehicles on the freeway had to wait for others to change the lane which in turn started to form queues. Furthermore, we defined the junction between the end of ramp lane and free way as "zipper node". A zipper node is a junction type where vehicles in the freeway get priority over vehicles at merging lanes and similar to real life the incoming vehicles wait till the freeway lane is clear. This modification made the simulation consistent with real-world freeway traffic and we noticed a better traffic flow than previous configuration. For the exit ramps, we also increased the length of the exit lane so that vehicles have more time to change lane and avoid last minute lane changes. In simulation, each segment of the road is called an edge and a junction or node connects two edges. To replicate free flow of traffic in freeways we made sure that the full network is continuous, and each junction is defined so that vehicles do not slow down or stop from going one edge to another.

__________________________________________________________________________________________________

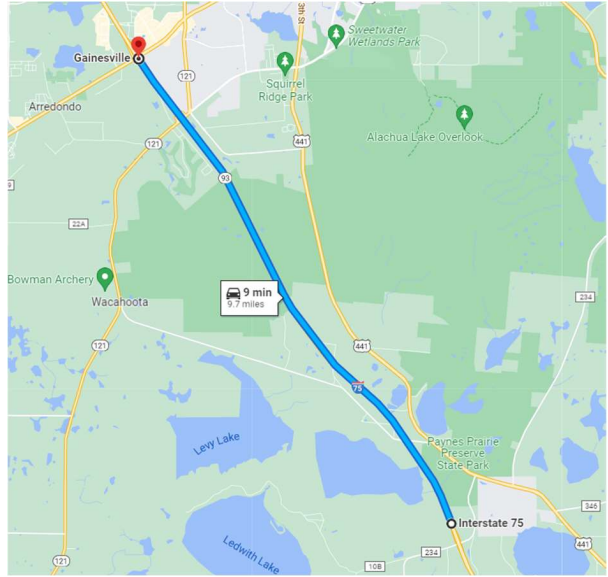

Figure 1: Road segments in I-75 used for simulation.

**Measures of Traffic and Potential Conflicts**

We assessed the traffic safety aspect during hurricane evacuation, along with average traffic flow and travel time for the selected road segments. To identify whether a vehicle is in a conflict when an intervention is required, a surrogate safety measure (SSM) is used (*29*). SSMs define traffic events that are deemed as possibly hazardous to participating road users. For this purpose, Time-to-collision (TTC) was chosen as an SSM in this research. Time-to-collision (TTC) is the primary conflict severity indicator, where a lower TTC value indicates a higher severity of a crash (*25*). TTC is defined as the expected time for two vehicles to collide, provided they both continue with their present speed and trajectory (*26*). TTC values are calculated by extrapolating vehicles trajectories, assuming constant velocity and unchanged course of collision (*27*).

When calculating TTC, a threshold needs to be selected. Any event where the TTC value is less than the threshold value the event is considered as a potential conflict. Previously, TTC was used with 1.5s threshold value for human-driven vehicles. In this experiment, we used a 0.5s threshold value for AV/CAVs and 1.5s for human-driven vehicles according to previous literatures (*4*),(*11, 16*). Since AV/CAVs have better reaction times, they should have a lower TTC threshold value (*17*).

**Simulation Parameters**

To simulate human-driven vehicles (HDVs), we used the same model calibrated in Rahman et al. (2021) (*4*) with same flow rate to replicate hurricane evacuation traffic. All the model parameters are shown in **Table 1**. The maximum speed was set at 70 mph which was the speed limit found in the interstate segment simulated.

The total simulation run time was 7200 seconds (2 hours) where the first 1800 seconds (30 minutes) and the last 1800 seconds data were considered as warm up and cool down periods, respectively and all results were collected from the middle 3600 seconds (1 hour). Previously 8 MVDS (Microwave Vehicle Detection System) detector data along the road segment were used to calibrate the network. The raw data, reported in every 20-30 seconds, were aggregated into 5-minutes intervals. The simulation network also contains loop detectors located near the same positions where the MVDS detectors were in the highway. We calculated the mean flow from these 8 loop detectors. Rahman et al. (2021) (*4*) used the Geoggrey E. Heaver statistics and chi-square statistics to compare the field volume with the simulated volume as part of the calibration process to find the evacuation traffic demand which were also used in this study.

When running the simulation experiments, we used a step length of 0.1s and changed the default lane changing model parameters. With a step length of 0.1s the simulation makes 10 times more calculations compared to a simulation performed with the default 1s step length. We observed that near the freeway ramps, due to heavy traffic, the number of potential conflicts increased, and queues were formed. A plausible explanation of these queues would be the lane changing behavior of the vehicles near ramps; this may also increase the number of potential conflicts. Thus, we explored different lane changing parameters in the simulation. In our experiments, we used the lane changing model LC2013 implemented in SUMO after modifying two default lane changing parameters: (i) "Strategic" (ii) "Cooperative". The "strategic" parameter is defined as the eagerness for performing strategic lane changing. Higher values result in earlier lane changing. The "co-operative" parameter is defined as the willingness for performing cooperative lane changing. Lower values result in reduced cooperation.

Human-driven vehicles (HDVs) may not always strategically change lanes beforehand which may result in more last-minute changes, especially when taking an exit. On the other hand, autonomous vehicles (AV/CAVs) with a set route are more likely to be strategic when changing lanes. Thus, we changed the "Strategic" parameter for AVs and CAVs to 100, as suggested in previous literature (*11*). This means AVs and CAVs will be able to anticipate lane changes 10 times earlier than HDV. In addition, a HDV is less likely to fully cooperate when changing lanes, thus we reduced the value of the lane changing parameter known as "Cooperative" from the default value of 1 to 0.8.

**Table 1: Simulation parameters for HDVs, AVs, and CAVs**

| Vehicle Type | Car Following Model | Max Speed (mph) | Speed Factor norm (mean, deviation, min, max) | Min Gap (m) | Max Accel. ($m/s^2$) | Max Decel. ($m/s^2$) | Desired Headway (s) |
|---|---|---|---|---|---|---|---|
| HDV | Krauss | 70 | normc(0.96,0.3,0.2,1) | 2 | 4.5 | 6.5 | 1.2 |
| CAV | CACC | 70 | normc(0.96,0.3,0.2,1) | 2 | 4.5 | 6.5 | 1.3 |
| AV | ACC | 70 | normc(0.96,0.3,0.2,1) | 2 | 4.5 | 6.5 | 1.3 |

**Simulating the effects of CAVs and AVs with SUMO**

To simulate the effects of CAVs, previous studies adopted the CACC car following model in SUMO. Through different experiments they found that CAVs can achieve higher traffic flow with better stability leading to improved traffic safety. However, we wanted to assess the feasibility and potential safety impacts of CAVs in our setting (evacuation traffic). We used different percentage of CAVs along with HDVs. Initially, we used the same traffic parameters as defined for the ACC parameters in Rahman et. al. (*4*) so that our results are comparable to previous literatures and our base case. We have changed only the desired headway of CAVs to 1.3s since AV/CAVs are likely to maintain a longer safe distance than human-driven vehicles.

To simulate the effects of AVs, we selected the ACC car following model available in SUMO and used the parameters reported in our previous study (*4*) which evaluated the effects of ACC equipped vehicles on evacuation safety. We ran the ACC car following model using the same traffic parameters (e.g., max speed, min gap, desired headway) used in CACC simulation. The model parameters for AVs and CAVs are given in Table 1.

The car-following models are empirically calibrated using real-world data by changing the values of the gain parameters ($k$). There are three gain parameters: $k_1, k_5, k_6$; the main difference from one operational mode of CAV to another depends on the value of these gain parameters. However, it is more challenging to set the gain parameters of CACC vehicles as we found a few discrepancies among the values previously used. Since we have no CACC data to calibrate the model, we relied on previous literature and ran a sensitivity analysis on three sets of values of the gain parameters, summarized in Table 2. The literature parameters consist of the values used by previous studies (13) (28) and the SUMO default parameters were the parameters set as default in the SUMO software. ACC vehicles rely only on on-board sensors to maintain a constant headway and does not have any vehicle-to-vehicle communication. Since ACC and CACC are inherently similar except the connectivity part, we set the corresponding ACC gain parameters to find the effects of those parameters on simulation results.

**Table 2: CACC gain parameters used in the simulation.**

| Gain Parameters | Parameters equivalent to ACC model | Literature parameters | SUMO defaults |
|---|---|---|---|
| speedControlGain, $k_1$ | -0.4 | -0.4 | -0.4 |
| gapClosingControlGainSpeed, $k_6$ | 0.8 | 1.6 | 0.05 |
| gapClosingControlGainSpace, $k_5$ | 0.04 | 0.01 | 0.005 |
| gapControlGainSpeed, $k_6$ | 0.07 | 0.25 | 0.45 |
| gapControlGainSpace, $k_5$ | 0.23 | 0.45 | 0.0125 |
| collisionAvoidanceGainSpace $k_6$ | 0.8 | 0.8 | 0.45 |
| collisionAvoidanceGainSpeed, $k_5$ | 0.23 | 0.23 | 0.05 |

**Co-Simulation of CAVs with SUMO and OMNET++**

In this research, a co-simulation framework, called Veins (Vehicles in Network Simulation), was used which integrates two distinct simulation platforms: SUMO, a microscopic traffic simulation model, and OMNeT++, a discrete event-based vehicular communication network simulation platform. This framework facilitates a reciprocal exchange of outputs between the two simulation tools, enabling a comprehensive simulation of traffic and communication network parameters without the necessity of introducing additional assumptions in either model. The study investigated various market penetration rates for Connected and Autonomous Vehicles (CAVs) and conducted a comparative analysis. Different market penetration rates for CAVs were studied and compared.

As an event-based simulator, OMNeT++ schedules node movements at regular intervals to handle mobility, which aligns well with SUMO's approach of advancing simulation time in discrete steps. The integrated control modules within OMNeT++ and SUMO effectively buffers incoming commands between timesteps to ensure synchronous execution at predefined intervals. During each timestep, OMNeT++ transmitted all buffered commands to SUMO and triggered the corresponding timestep in the road traffic simulation. After completion of the road traffic simulation timestep, SUMO sent back a series of commands along with the positions of all instantiated vehicles to the OMNeT++ module.

In OMNeT++ each CAV is considered as a mobile node. The two-way communication between the two simulators allowed OMNeT++ to respond to the received mobility trace by introducing new nodes, removing nodes that had reached their destinations, and adjusting the positions of nodes based on their road traffic simulation counterparts. Subsequently, OMNeT++ advanced the simulation until the next scheduled timestep, enabling nodes to adapt to altered environmental conditions, influenced by the Inter-Vehicle Communication (IVC), which impacted their speed and routes.

The purpose of this study is to study the effect of CAV with realistic traffic and communication parameters. To achieve this, we implemented a straightforward warning strategy for the CAVs. When a vehicle comes to a complete stop (with a speed of zero), it waits for a certain duration (approximately 1 second in our simulations) before alerting other vehicles about a potential incident on the current lane. If the network simulation confirms that another vehicle eventually receives this incident warning, it records the timestamp and content of the warning message. This allows participating vehicles in the Inter-Vehicle Communication (IVC) to adjust their routes and avoid the incident. As our network consists of only one segment, most vehicles have the option to change lanes or closely follow the leading vehicle. Once the originating vehicle resumes its journey, it notifies other vehicles that the lane is safe to use again, enabling them to revert to their original estimated travel times.

In the simulation, the networking parameters are set up to closely mimic real-world conditions. The complete network stack, including Address Resolution Protocol (ARP), is simulated, and the wireless modules are configured to resemble IEEE 802.11b network cards transmitting at 6 Mbps with Request to Send (RTS) and Clear to Send (CTS) disabled. To model radio wave propagation, a plain free-space model is used. These parameters ensure a comprehensive representation of communication behavior in the simulation, facilitating a thorough study of the impact of the network on the Connected and Autonomous Vehicles (CAVs) scenario being investigated.

**RESULTS**

We calibrated the Krauss model (base case) with modified lane change parameters, the ACC model, and CACC model with proper gain and lane changing parameters. Lists of the experimented and selected parameters are given in Table 3. All the scenarios were run with these parameters along with the traffic parameters given in Table 1. After finalizing all the parameters, we ran each scenario ten times and tabulated all the results in Table 4 and Table 5.

**Table 3: Summary of the parameters selected from experiments.**

| Type of Parameter | Parameter | Experimental values used | Selected value |
|---|---|---|---|

Syed and Hasan| Lane changing parameter | Strategic lane change | HDV= [1, 10] AV/CAV= [1, 100] | HDV=10 AV/CAV=100 |
|---|---|---|---|
| | Co-operative lane change | HDV = [1, 0.8] | HDV=0.8 |
| Model Parameter | CACC gain parameter sets | [Literature parameter, Parameter equivalent to ACC, SUMO defaults] | Literature parameters |
| Traffic Parameters | Speed deviation | CACC= [0.1, 0.05] | 0.05 |

Our results show a 65.9% drop in potential conflicts with 25% AV which is a 16.2% improvement from the previously reported result (*4*). This improvement in safety is mainly due to the modification of lane changing parameter as most of the other parameters are kept same. The change ensured AVs anticipate lane changes 10 times earlier and are more co-operative in changing lanes with respect to HDVs. The modifications in the network have also ensured more realistic traffic movement specially in the merging lanes and less ques formation which may have led to less number of potential conflicts. The ACC car following model has outperformed the CACC car following model at every corresponding scenario with a lower number of potential conflicts, less average travel time, and higher average traffic flow. The standard deviation values are also significantly less meaning that the ACC car following model produces more stable and consistent results in simulating the effects of AVs.

The results are highly fluctuating for CACC scenarios with very high standard deviation values at each scenario. We found the "literature parameters" to produce the best and relatively more stable results in comparison to the other parameter sets. This fluctuation indicates that the CACC car-following model, implemented in SUMO simulation model, is unstable in simulating the effects of CAVs for the chosen hurricane evacuation scenario. In addition to the unstable behavior of the CACC simulation results, we find that the number of potential conflicts in CACC-25 and CACC-50 scenarios are higher than the base case scenario. Then in the CACC-75 scenario, it significantly drops to a total 173 potential conflicts, a modest 6.2% drop from the base case scenario (Krauss-100).

The co-simulation results reveal that as the percentage of Connected and Autonomous Vehicles (CAVs) increases, the Time to Collision (TTC) decreases, indicating improved safety. We observed a 42.5% drop in potential conflict with 25% introduction of CAVs from our co-simulation study. The trend does not follow linearly as we see that a decrease of TTC by 56% and 53% for CAV-50 and CAV-75 scenarios, respectively. While traffic flow is lower compared to the base case, the deviation is at most 2.8%, with the lowest traffic flow occurring at CAV-25 and increasing with higher CAV percentages. However, travel time also increases with more CAVs due to adjustments in traffic patterns and behavior. Nevertheless, CAVs show higher stability and reliability than simulations with Cooperative Adaptive Cruise Control (CACC) models, making them a promising technology for enhancing transportation systems with improved safety, traffic flow, and reliability.

Syed and Hasan

**Table 4: Number of potential conflicts for all scenarios using the final parameters**

| Scenario | HDV TTC count | std for TTC-HDV | (C)AV TTC count | std for TTC-CACC | Total TTC count | std deviation | Percentage change from base case |
|---|---|---|---|---|---|---|---|
| Krauss-100 (base case) | 184.4 | 28.34 | 0 | 0 | 184.4 | 28.34 | 0 |
| ACC-25 | 62.8 | 24.07 | 0 | 0 | 62.8 | 24.07 | -65.9 |
| ACC-50 | 21.2 | 7.19 | 0 | 0 | 21.4 | 7.50 | -88.4 |
| ACC-75 | 8.8 | 5.07 | 0 | 0 | 8.8 | 5.07 | -95.2 |
| ACC-100 | 0 | 0 | 0 | 0 | 0 | 0 | -100 |
| CACC-25 | 328.6 | 259.74 | 15 | 14.09 | 343.6 | 273.57 | 86.3 |
| CACC-50 | 342.6 | 224.98 | 24.8 | 20.32 | 367.4 | 244.97 | 99.2 |
| CACC-75 | 161.6 | 49.88 | 11.4 | 1.52 | 173 | 49.90 | -6.2 |
| CACC-100 | 0 | 0 | 0.2 | 0.45 | 0.2 | 0.45 | -99.9 |
| CAV-25 | 106 | 0 | 106 | 0 | 106 | 0 | -42.5 |
| CAV-50 | 80 | 0 | 80 | 0 | 80 | 0 | -56.6 |
| CAV-75 | 86 | 0 | 86 | 0 | 86 | 0 | -53.4 |
| CAV-100 | 0 | 0 | 0 | 0 | 0 | 0 | -100 |

* CAV-XX represents the co-simulation scenarios representing actual CAV and XX denotes the corresponding percentage.

**Table 5: Traffic measures for all scenarios using the final parameters.**

| Scenario | Average travel time (minutes) | Percentage change from base case | Average traffic flow (vehicle per hour) | Percentage change from base case |
|---|---|---|---|---|
| Krauss-100 | 10.154 | 0 | 4495.47 | 0 |
| ACC-25 | 9.995 | -1.6 | 4503.76 | 0.2 |
| ACC-50 | 9.99 | -1.6 | 4504.68 | 0.2 |
| ACC-75 | 9.98 | -1.7 | 4522.06 | 0.6 |
| ACC-100 | 9.97 | -1.8 | 4556.03 | 1.35 |
| CACC-25 | 10.71 | 5.4 | 4491.89 | -0.1 |
| CACC-50 | 12.80 | 26.1 | 4467.73 | -0.6 |
| CACC-75 | 12.31 | 21.2 | 4469.82 | -0.6 |
| CACC-100 | 12.86 | 26.67 | 4391.39 | -2.32 |
| CAV-25 | 10.64 | 4.83 | 4368.15 | -2.83 |
| CAV-50 | 10.96 | 7.97 | 4383.56 | -2.49 |
| CAV-75 | 12.01 | 18.28 | 4440.50 | -1.22 |
| CAV-100 | 12.21 | 20.2 | 4410.11 | -1.90 |

## CONCLUSIONS

Overall, the simulation experiments conducted in this study provides insights on improving the simulation-based approaches to assess different vehicle technology in a hurricane evacuation scenario. Despite many positive results found in the literature for the CACC car-following models representing CAVs, our results suggest that the CACC car-following model produces very unstable results reflected by high standard deviations in reported results. From the unstable nature of the result and lack of real-world CAV data it is unclear whether the CACC car-following model should be used to model CAVs. Our results suggest that the CACC car following model is not appropriate to simulate the effects of CAVs, at least, in a hurricane evacuation context. However, we confirmed and enhanced the previously found results that ACC equipped vehicles (equivalent to AVs), even with a 25 % market penetration rate, can significantly reduce the number of conflicts. With the modified lane changing parameters we showed that, with only a 25% market penetration rate, AVs can decrease the number of potential conflicts by 65.9% (vs. 49.7% found in our previous study). Our results also suggest that the ACC car following model is more stable and performing better than the CACC car following model in terms of traffic safety and traffic flow. Thus, we may continue to simulate the effects AVs with ACC car following model.

Additionally, this research explores the impact of Connected and Autonomous Vehicles (CAVs) on evacuation traffic dynamics and safety through co-simulation experiments with varying percentages of CAVs. The results reveal significant improvements in safety, as evidenced by the 42.5% decrease in potential number of conflicts with respect to base case scenario for 25% of CAVs. While traffic flow decreases compared to the base case scenario, it remains stable with a maximum deviation of 2.8%. Travel time increases as the percentage of CAVs rises, but the overall results demonstrate enhanced stability and reliability in comparison to simulations using Cooperative Adaptive Cruise Control (CACC) car following models. These findings highlight the potential benefits of integrating CAVs into traffic systems, paving the way for safer and more efficient transportation networks in the future.

## ACKNOWLEDGMENTS

The authors are grateful to the US National Science Foundation for grants 1917019 and 2122135 and SAFER-SIM University Transportation Center to support the research presented in this paper. However, the authors are solely responsible for the findings presented here.

## AUTHOR CONTRIBUTIONS

The authors confirm contribution to the paper as follows: study conception and design: Syed and Hasan; data collection: Syed; simulation experiments, data analysis and interpretation of results: Syed; draft manuscript preparation and editing: Syed and Hasan; and funding acquisition and supervision: Hasan. All authors reviewed the results and approved the final version of the manuscript.